\documentstyle[11pt,adassconf]{article}  
\begin{document}   

\paperID{P4.1}

\title{Pointing Refinement of SIRTF Images}

\author{F. Masci, D. Makovoz, M. Moshir, D. Shupe and John W. Fowler}
\affil{SIRTF Science Center, California Institute of Technology,
       Pasadena, CA 91125, Email: fmasci@ipac.caltech.edu}

\contact{Frank Masci}
\email{fmasci@ipac.caltech.edu}

\paindex{Masci, F.}
\aindex{Makovoz, D.}
\aindex{Moshir, M.}
\aindex{Shupe, D.}
\aindex{Fowler, J.}


\begin{abstract}          
The soon-to-be-launched Space Infrared Telescope Facility (SIRTF) shall 
produce image data with an a-posteriori pointing knowledge of $1.4\arcsec$ 
($1\sigma$ radial) with a goal of $1.2\arcsec$ in the International Celestial Reference System (ICRS).
To perform robust image coaddition, mosaic generation,
extraction and position determination of faint sources,
the pointing will need to be refined to better than a few-tenths of an arcsecond.
This paper summarizes the pointing-refinement algorithm and presents
the results of testing on simulated data.
\end{abstract}

\section{Introduction}

One of the goals of astronomical image data acquisition is to infer the
pointing in an absolute coordinate system as accurately as possible, 
in this case the celestial reference system. Instabilities 
in telescope pointing and tracking however inhibit us from achieving this goal.
SIRTF for instance, is expected to provide 
pointing and control of at least $5\arcsec$ absolute accuracy with 
$0.3\arcsec$ stability over 200 sec ($1\sigma$ radial). 
In the ICRS, the star-tracker assembly provides 
an a-posteriori pointing knowledge of $1.4\arcsec$.
The end-to-end pointing 
accuracy is a function of the inherent star-tracker accuracy, 
the spacecraft control system, how well the 
star-tracker bore-sight is known in the focal plane array (instrument) frame, 
and variations in the latter due to thermo-mechanical deflections.
The SIRTF observatory will have a data flow rate of $> 10,000$ 
science images per day. This will require an automated, self-consistent means
of refining the celestial pointing as robustly as possible.

The conventional method to refine the pointing is to
make comparisons of astrometric sources with positions known to 
better than a few percent of the observed positional errors.
The primary motivation for pointing refinement is to enable robust coaddition of
image frames in a common reference frame so that source extraction and position
determination to faint flux levels can be performed. Moreover, refinement in
an absolute (celestial) reference frame will enable robust cross-identification
of extracted sources with other catalogs. 

For these purposes, we have developed a 
stand-alone software package ``{\it pointingrefine}'' with a goal to 
generate science products with sub-arcsecond pointing accuracy in the ICRS. 
The software can refine the pointings and orientations of SIRTF images in either
a ``relative'' sense where pointings become fixed relative to a single image
of a mosaic, or, in an ``absolute'' sense (in the ICRS) when absolute point source
information is known.

\section{Software Summary}

As part of routine pipeline operations at the SIRTF Science Center (SSC),  
all input images are pre-processed for
instrument artifact removal and pointing data attached to raw FITS images.
Following this, point source extraction
is performed on individual frames for input into {\it pointingrefine}.
The software expects point source lists adhering to the format produced
by the SSC source extractor.   
The {\it pointingrefine} software performs the following:  

\begin{enumerate}
\item Reads source extraction tables, FITS images from input lists and if 
      absolute refinement is desired, a list of absolute point source positions.
\item Point source position and flux matching is performed between all possible
      image pairs in the input list. This includes absolute point sources if available.
\item Transform correlated point source positions
      and uncertainties to a Cartesian fiducial (mosaic) reference frame.
\item Set up global minimization equations involving relative offsets between
      all correlated point source positions.
\item Solve a linear matrix equation for translational and rotational offsets in
      mosaic reference frame for each input image and compute the full
      error-covariance matrix.
\item Apply these offsets to the pointing centers ($x_{c}$, $y_{c}$) of all images. 
      Transform to RA, Dec to obtain refined celestial pointings and 
      orientations.
\item Results output: Write new refined pointing keywords to FITS headers and
      to a table file with diagnostic information.
\end{enumerate}

\section{Algorithm}

A brief outline of the refinement algorithm (primarily steps 4, 5, and 6 above)
is as follows.
\begin{figure}
\epsscale{0.7}
\plotone{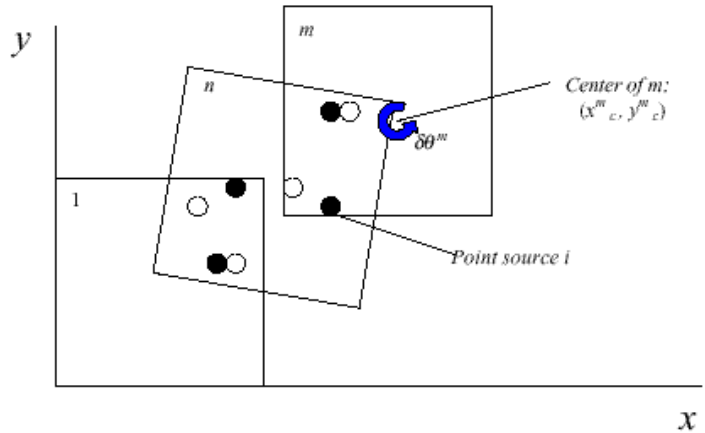}
\vspace{-5mm}
\caption{A simple three image mosaic.}\label{fig1}
\end{figure}
Consider the simple three image mosaic in Figure~\ref{fig1}. Image ``1''
defines the ``fiducial'' reference frame. The circles represent point sources
detected from each overlapping image pair transformed into the fiducial frame. 
The filled circles are sources extracted from image {\it n} and the open
circles are sources extracted from either image 1 or {\it m}. The correlated
source pairs are slightly
offset from each other to mimic the presence of pointing uncertainty in each
raw input image. The Cartesian coordinates of a correlated point source \underline{common}
to an image pair are related by:

\begin{eqnarray}
x^m_i \rightarrow \tilde{x}^n_i & = & x^m_i - (y^m_i - y^m_c)\delta \theta^m + \delta X^m \nonumber\\ & & \\
y^m_i \rightarrow \tilde{y}^n_i & = & y^m_i - (x^m_i - x^m_c)\delta \theta^m + \delta Y^m , \nonumber
\end{eqnarray}
where $\delta\theta^m$, $\delta X^m$, $\delta Y^m$ are rotational and
Cartesian offsets respectively in the fiducial mosaic frame. We have
made the approximation $\sin\delta\theta\approx\delta\theta$ 
($\cos\delta\theta\approx 1$) since uncertainties in measured position angles are
expected to be small 
(\raisebox{-0.7ex}{\mbox{$\stackrel{\textstyle <}{\sim}$}} 20$\arcsec$).

We define a cost function $L$, representing the sum of the squares of the
``corrected'' differences of all correlated point source positions in all
overlapping image pairs ($m$,$n$):

\begin{equation}
L = \sum_{m,n}\sum_{i}\left\{\frac{1}{\Delta x_i^{m,n}}[\tilde{x}^n_i - \tilde{x}^m_i]^2 +
                        \frac{1}{\Delta y_i^{m,n}}[\tilde{y}^n_i - \tilde{y}^m_i]^2\right\} ,
\end{equation}
where
\begin{eqnarray}
\Delta x_i^{m,n} & = & \sigma^2(x^m_i) + \sigma^2(x^n_i)\\ \nonumber
\Delta y_i^{m,n} & = & \sigma^2(y^m_i) + \sigma^2(y^n_i)\nonumber
\end{eqnarray}
and the $\sigma^2$ represent variances in extracted point source positions. 
The function $L$ (Equation 2) is minimized with respect to the Cartesian
offsets ($\delta\theta^m$, $\delta X^m$, $\delta Y^m$) for an image $m$
which contains point sources in common with all other images $n$. At the global
minimum of $L$, the following conditions hold:

\begin{equation}
\frac{\partial L}{\partial\delta\theta^m} = 0;\;
\frac{\partial L}{\partial\delta X^m} = 0;\;
\frac{\partial L}{\partial\delta Y^m} = 0.
\end{equation}
Evaluating these partial derivatives leads to a set of three simultaneous
equations for each image in our mosaic. In general, for $N$ correlated images,
we will have $3(N-1)$ simultaneous equations in $3(N-1)$ unknowns. It is ``$N-1$''
because we exclude the reference image frame which by definition has 
the constraint  
$\delta\theta^m = \delta X^m = \delta Y^m = 0$. The $3(N-1)$ system of equations can 
be solved using linear sparse matrix methods since a large number of the matrix 
elements will be zero. We use the
\htmladdnormallinkfoot{UMFPACK}{http://www.cise.ufl.edu/research/sparse/umfpack/}
library which is adapted for solving large unsymmetric matrix systems.

When absolute astrometric references are available,
the ``fiducial'' reference image (mosaic) frame is treated like a single input image,
which contains the {\it absolute} point source positions. When the input images
become refined relative to this fiducial image, they in reality become {\it absolutely}
refined (in the ICRS). The presence of absolute point sources also reduces the effect of ``random walks''
in offset uncertainties with distance if a single input image were chosen as the
reference instead. Once Cartesian offsets ($\delta\theta^m$, $\delta X^m$, $\delta Y^m$)
in the reference (mosaic) frame are computed, the
pointing centers ($x_{c}$, $y_{c}$) are corrected (by use of Equation 1) and 
transformed back to the sky 
to yield refined pointings. Image orientations are refined in a similar manner, but in this
case, we need to transform at least two fiducial points per image to uniquely determine
the orientation.

\section{Simulations}

The Infrared Array Camera (IRAC) on SIRTF will perform simultaneous imaging at four
bands spanning the range $\approx 3.6\mu m$ to $8\mu m$. Each array
consists of $256\times 256$ $1.2\arcsec$ pixels. We simulated a mosaic
of 800 IRAC ($3.6\mu m$) ``truth'' images (i.e. {\it with no} pointing
error) with each image containing randomly distributed point sources. Input image overlap 
coverage was $\sim 50\%$. A second set of 800 images was
simulated with random errors added to the pointing keywords of
image headers. The errors were drawn from a Gaussian distribution with
mean radial error $\langle\delta\rangle\approx1.4\arcsec$. An absolute point source
list was also simulated by extracting the brightest sources with smallest centroiding errors
from each ``truth'' image. The average number of ``absolutes'' per input image
was 10. The SSC point source extractor was then used to extract $\approx 50$ point sources
from each input image (with pointing error).

\begin{figure}
\epsscale{0.65}
\vspace{-10mm}
\plotone{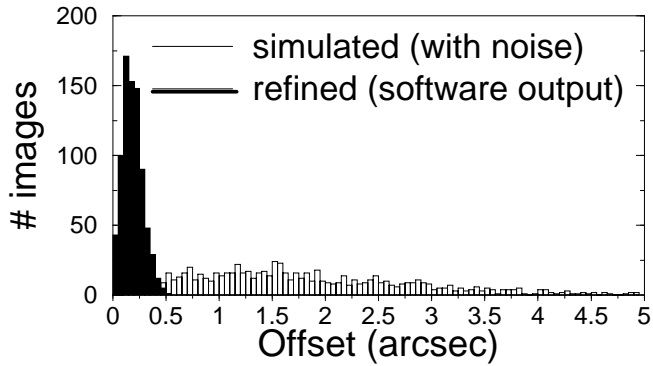}
\vspace{-6mm}
\caption{Distribution of radial offsets between ``true'' (noiseless) 
and simulated pointings (with noise) - {\it broad histogram}, 
and between true and refined pointings - {\it narrow histogram}.}\label{fig2}
\end{figure}

Figure~\ref{fig2} shows the results of our 
simulation where we compare the distributions of radial offsets relative to ``truth'' 
pointings before and after refinement. The refinement is better than
$85\%$ for almost every image. The main limitation is full knowledge of the
Point Response Function (PRF)
to reduce centroiding errors in source extractions. However, we expect extraction centroids better
than $0.1\arcsec$ with better sampled PRFs. This will give us the 
sub-arcsecond absolute pointing accuracy sought in SIRTF's imaging detectors. 

\acknowledgments
This work was carried out at the SIRTF Science Center, with 
funding from NASA under contract to the California Institute of
Technology and the Jet Propulsion Laboratory.
 
\end{document}